\documentclass[conference]{IEEEtran}
\usepackage{amsmath,amsthm,amssymb}
\usepackage{graphicx}
\usepackage{stmaryrd}
\usepackage{blkarray}

\theoremstyle{plain}

\newtheorem{theorem}{Theorem}

\DeclareMathOperator{\F}{\mathbb F}
\DeclareMathOperator{\W}{\mathcal W}

\newcommand{\mF}{\mathcal F}
\newcommand{\set}[1]{\left\{{#1}\right\}}

\DeclareMathOperator{\diag}{diag}

\DeclareMathOperator{\sgn}{sgn}

\begin{document}

\title{Chained  Successive Cancellation Decoding of the Extended Golay code}
\author{Peter Trifonov}
\author{\IEEEauthorblockN{Peter Trifonov} 
\IEEEauthorblockA{Saint Petersburg Polytechnic University,
Russia\\Email: petert@dcn.icc.spbstu.ru}}

\maketitle

\begin{abstract}
    The extended Golay code is shown to be representable as a chained polar subcode. This enables its decoding with the successive cancellation decoding algorithm and its stack generalization. The decoder can be further simplified by employing fast Hadamard transform. The complexity of the obtained algorithm is comparable with that of the Vardy algorithm. 
\end{abstract}

\section{Introduction}
The $(24,12,8)$ extended Golay code is a quasi-perfect self-dual linear binary block code. It has found numerous applications in communication, storage and imaging systems \cite{golay1949notes,garcia2010application,Hussain2008Golay,key1999some}.
Rich algebraic structure of the Golay code admits very efficient decoding, see \cite{vardy1995even} and references therein. However, these algorithms are specific to the extended Golay code, and, in general, may not be used for decoding of other types of error correcting codes. 

Polar codes is a novel class of capacity-achieving error correcting codes, which have very efficient construction, encoding and decoding algorithms \cite{arikan2009channel}. Furthermore, the list and sequential successive cancellation decoding algorithms \cite{tal2015list,miloslavskaya2014sequential} were shown to be applicable for decoding of short extended BCH\ codes \cite{trifonov2016polar}.
Polar codes were adopted for use in 5G wireless, so many future communication systems are likely to have an implementation of a decoder for polar codes. 
It is tempting to explore application of the  decoding techniques developed for polar codes for other types of error correcting codes. This would enable communication systems to support different channel coding schemes with the same hardware.

In this paper we show that the extended Golay code can be represented in the framework of chained polar subcodes  \cite{trifonov2017chained}, and suggest a low-complexity decoding algorithm based on this representation. The proposed algorithm can be considered as a generalization of sequential (stack) and block sequential decoding algorithms \cite{miloslavskaya2014sequential,niu2012stack,trofimiuk2015block}.

The paper is organized as follows. In section \ref{sBG} we review polar codes, their generalizations and decoding algorithms. Section \ref{seGolay} introduces a  representation of the extended Golay code as a chained polar subcode.
This representation is used in Section \ref{sDecoding} to derive some new decoding algorithms. Simulation results are presented in Section \ref{sNumeric}.  
\section{Background}
\label{sBG}
\subsection{Dynamic frozen symbols}
$(n=2^m,k)$ polar code is a set of vectors $c_0^{n-1}=u_0^{n-1}A_m$,
where $A_m=B_m\begin{pmatrix}1&0\\1&1\end{pmatrix}^{\otimes m}$ is the polarizing matrix, $u_i=0,i\in \mF$, $B_m$ is the bit reversal permutation matrix, and $\mF\subset{0,\dots,2^m-1}$ is the set of $2^m-k$ frozen channel indices \cite{arikan2009channel}. 
It is possible to show that $A_m$ together with a binary input memoryless channel $\mathbf W(y|c)$ give rise to synthetic bit subchannels 
$$\mathbf W_m^{(i)}(y_0^{n-1},u_0^{i-1}|u_i)=\frac{1}{2^{n-1}}\sum_{u_{i+1}^{n-1}}\prod_{j=0}^{n-1}\mathbf W(y_j|(u_0^{n-1}A_{m})_j),$$
and the capacities of these subchannels converge with $m$ to 0 or 1 bits per channel use. The standard way to construct polar codes is to let $\mF$
be the set of low-capacity subchannels.
However, the minimum distance of classical polar codes is quite low. It was suggested in \cite{trifonov2016polar} to select $u_0^{n-1}$ in such way, so that the obtained vectors $c_0^{n-1}$ are codewords of some linear block code with check matrix $H$. This corresponds to dynamic freezing constraints \begin{equation}
\label{mDynFrozen}
u_i=\sum_{j<i}V_{s_i,j}u_j,i\in \mF,
\end{equation}
where $V=QHA_m^T$ is a $(n-k)\times n$ constraint matrix, and $Q$ is an invertible matrix, such that last non-zero elements in rows of $V$ are located in distinct columns,  $\mF$ is the set of indices of such columns, and $s_i$ is the index of the row having the last non-zero entry in column $i$. Alternatively, the codewords of a polar subcode can be obtained as $c_0^{n-1}=xWA_m$, where $W$ is a $k\times n$ precoding matrix, such that $WV^T=0$, and $x$ is an information vector.

 Decoding of such codes can  be implemented with a straightforward generalization of the successive cancellation decoding algorithm, which makes decisions
\begin{equation}
\label{mSC}
\hat u_i=\begin{cases}
\arg \max_{u_i\in \F_2}\mathbf W_m^{(i)}(y_0^{n-1},\hat u_0^{i-1}|u_i),&i\notin \mF\\
\sum_{j<i}V_{s_i,j}\hat u_j,&i\in \mF.
\end{cases}
\end{equation}
Extended primitive narrow-sense BCH\ codes were shown to have particularly well-structured sets of frozen symbol indices, and admit efficient list/sequential SC decoding  \cite{trifonov2016polar}.

Representation of linear codes via the dynamic freezing constraints can be considered as a result of application of the generalized Plotkin decomposition introduced in \cite{trifonov2016polar}.
\begin{theorem}[\cite{trifonov2016polar}]
Any  linear $(2n,k,d)$ code $\mathcal C$ has a generator matrix given by 
\begin{equation}
\label{mGenPlotkin}
G=\begin{pmatrix}
I_{k_1}&0&\tilde I \\
0& I_{k_2}&0\\
\end{pmatrix}\begin{pmatrix}
G_1&0\\
G_2&G_2\\
G_3&G_3\\
\end{pmatrix},
\end{equation}
where $I_l$ is a $l\times l$ identity matrix,  $G_i, 1\leq i\leq 3,$   are  $k_i\times n$ matrices,  $k=k_1+k_2$,  and $\tilde I$ is obtained by stacking a $(k_1-k_3)\times k_3$ zero matrix and $I_{k_3}$ , where $k_3\leq k_1$.
\end{theorem}
In this paper we essentially present a generalization of  this decomposition.

\subsection{List successive cancellation decoding}
In general, classical polar codes, polar subcodes and other codes represented by \eqref{mDynFrozen} require list successive cancellation decoding in order to obtain near-ML\ performance. 
Let $$\W_m^{(i)}(u_0^i|y_0^{n-1})=\max_{u_{i+1}^{n-1}\in\F_2^{n-i-1}}\mathbf W_m^{(n-1)}(u_0^{n-1}|y_0^{n-1})$$
be the probability of the most likely continuation of path $u_0^i$ in the code tree, without taking into account freezing constraints on symbols $u_j,j>i$.
It can be seen that for $\lambda>0$ 
\begin{align}
\W_\lambda^{(2i)}(u_0^{2i}|y_0^{n-1})=&\max_{u_{2i+1}\in\F_2}\W_{\lambda-1}^{(i)}(u_{0,e}^{2i+1}\oplus u_{0,o}^{2i+1}|y_0^{\frac{n}{2}-1})\nonumber\\
&\cdot \W_{\lambda-1}^{(i)}(u_{0,o}^{2i+1}|y_{\frac{n}{2}}^{n-1}),\\
\W_\lambda^{(2i+1)}(u_0^{2i+1}|y_0^{n-1})=&\W_{\lambda-1}^{(i)}(u_{0,e}^{2i+1}\oplus u_{0,o}^{2i+1}|y_0^{\frac{n}{2}-1})\nonumber\\
&\cdot \W_{\lambda-1}^{(i)}(u_{0,o}^{2i+1}|y_{\frac{n}{2}}^{n-1}),
\end{align}
and $\W_0^{(0)}(c|y_j)=\mathbf W(c|y_j)$. Let us define modified log-likelihood ratios $$S_\lambda^{(i)}(u_0^{i-1},y_0^{n-1})=\log\frac{\W_\lambda^{(i)}(u_0^{i-1}.0|y_0^{n-1})}{\W_\lambda^{(i)}(u_0^{i-1}.1|y_0^{n-1})}.$$
It is possible to show  that \cite{miloslavskaya2014sequential,trifonov2017star}
\begin{align*}
S_{\lambda}^{(2i)}(u_0^{2i-1},y_0^{N-1})=&a\boxplus b= \sgn (a)\sgn (b)\min(|a|,|b|)\\
S_{\lambda}^{(2i+1)}(u_0^{2i},y_0^{N-1})=&(-1)^{u_{2i}}a+b,
\end{align*}
where $a=S_{\lambda-1}^{(i)}(u_{0,e}^{2i-1}\oplus u_{0,o}^{2i-1},y_0^{\frac{N}{2}-1})$, $b=S_{\lambda-1}^{(i)}(u_{0,o}^{2i-1},y_{\frac{N}{2}}^{N-1})$, $N=2^\lambda$.  Then the logarithm of the probability of the most likely continuation of a path $u_0^i$ can be obtained as
\begin{align}
R(u_0^i|y_0^{n-1})=&\log\W_m^{(i)}(u_0^i|y_0^{n-1})\nonumber\\
=&R(u_0^{i-1}|y_0^{n-1})+\tau\left(S_m^{(i)}(u_0^{i-1},y_0^{n-1}),u_i\right),\label{mScoreMinSum}
\end{align}
where  $$\tau(S,u)=\begin{cases}
0,&\sgn(S)=(-1)^u\\
-|S|,&\text{otherwise.}
\end{cases}$$
One can assume that $R(\epsilon|y_0^{n-1})=0$, where $\epsilon$  is an empty sequence.
Observe that $R(u_0^i|y_0^{n-1})$ is equal up to the sign to the approximate path metric introduced in \cite{balatsoukasstimming2015llrbased}. The above derivation shows that this value is not just an approximation to the path metric used by the Tal-Vardy list decoder, but  reflects the likelihood of the most probable continuation of a path in the code tree, without taking into account not-yet-processed freezing constraints. 

It can be also seen that $R(u_0^{n-1}|y_0^{n-1})=-E(u_0^{n-1}A_m,y_0^{n-1})$, where 
$$E(c_0^{n-1},y_0^{n-1})=-\sum_{j=0}^{n-1}\tau(S_0^{(0)}(y_i),c_i)$$
is the ellipsoidal weight or correlation discrepancy of vector $c_0^{n-1}$ with respect to the noisy vector $y_0^{n-1}$.
\subsection{Chained polar subcodes}
Classical polar codes are limited to length $2^m$. In order to obtain codes of arbitrary length, it was suggested in \cite{trifonov2017chained} to 
combine polarizing matrices of different size. That is, the codewords of chained polar subcodes are given by  $c_0^{n-1}=xW\underbrace{\diag(A_{m_0},\dots,A_{m_{s-1}})}_A$, where $n=\sum_{i=0}^{s-1}2^{m_i}$, and $A$ is the mixed polarizing transformation matrix. A generalization of the successive cancellation decoding algorithm and its derivatives to the case of chained polar subcodes is provided in \cite{trifonov2017chained}. Alternatively, the code can be described as a set of vectors $c_0^{n-1}=u_0^{n-1}A$, where $u_0^{n-1}V^T=0$, and $V$ is the constraint matrix, such that $WV^T=0$.

In general, list or sequential decoding algorithm should be used for decoding of chained polar subcodes. These algorithms essentially operate by arranging the input symbols of polarizing transformations $A_{m_i}$ in some order, called decoding schedule,  and interleaving steps of conventional list/sequential successive cancellation for each $A_{m_i}$. The performance of such algorithm does depend on  the ordering of symbols $u_i$. It was shown in \cite{trifonov2017chained}  that the best performance is achieved by the greedy schedule, which aims on processing of frozen symbols as early as possible.

\section{The extended Golay code}
\label{seGolay}
\begin{figure*}[!t]
$$G=H=\left(\begin{array}{cccccccc|cccccccc|cccccccc}
1&1&1&1&0&0&0&0&0&0&0&0&0&0&0&0&1&1&1&1&0&0&0&0\\
1&0&1&0&1&0&1&0&0&0&0&0&0&0&0&0&1&0&1&0&1&0&1&0\\
1&0&0&1&1&0&0&1&0&0&0&0&0&0&0&0&1&0&0&1&1&0&0&1\\
1&0&0&1&0&1&1&0&0&0&0&0&0&0&0&0&1&0&0&1&0&1&1&0\\\hline
0&0&0&0&0&0&0&0&1&1&1&1&0&0&0&0&1&1&1&1&0&0&0&0\\
0&0&0&0&0&0&0&0&1&0&1&0&1&0&1&0&1&0&1&0&1&0&1&0\\
0&0&0&0&0&0&0&0&1&0&0&1&1&0&0&1&1&0&0&1&1&0&0&1\\
0&0&0&0&0&0&0&0&1&0&0&1&0&1&1&0&1&0&0&1&0&1&1&0\\\hline
1&1&0&1&1&0&0&0&1&1&0&1&1&0&0&0&1&1&0&1&1&0&0&0\\
1&0&1&1&0&0&1&0&1&0&1&1&0&0&1&0&1&0&1&1&0&0&1&0\\
1&0&0&0&1&0&1&1&1&0&0&0&1&0&1&1&1&0&0&0&1&0&1&1\\
1&0&0&1&0&1&0&1&1&0&0&1&0&1&0&1&1&0&0&1&0&1&0&1
\end{array}
\right)$$
$$V=QHA^{T}=\left(\begin{array}{cccccccccccccccc|cccccccc}
1&0&0&0&0&0&0&0&0&0&0&0&0&0&0&0&0&0&0&0&0&0&0&0\\
0&1&0&0&0&0&0&0&0&0&0&0&0&0&0&0&0&0&0&0&0&0&0&0\\
0&0&1&0&0&0&0&0&0&0&0&0&0&0&0&0&0&0&0&0&0&0&0&0\\
0&0&0&0&1&0&0&0&0&0&0&0&0&0&0&0&0&0&0&0&0&0&0&0\\
0&0&0&0&0&0&0&0&1&0&0&0&0&0&0&0&0&0&0&0&0&0&0&0\\
0&0&0&0&0&0&0&0&0&0&0&0&0&0&0&0&1&0&0&0&0&0&0&0\\
0&0&0&1&0&0&0&0&0&0&0&0&0&0&0&0&0&1&0&0&0&0&0&0\\
0&0&0&0&0&1&0&0&0&0&0&0&0&0&0&0&0&0&1&0&0&0&0&0\\
0&0&0&0&0&0&1&0&0&1&0&0&0&0&0&0&0&1&1&1&0&0&0&0\\
0&0&0&0&0&0&0&0&0&1&0&0&0&0&0&0&0&0&0&0&1&0&0&0\\
0&0&0&0&0&0&0&0&0&0&1&0&0&0&0&0&0&1&1&0&0&1&0&0\\
0&0&0&0&0&0&0&0&0&0&0&0&1&0&0&0&0&1&0&0&0&0&1&0
\end{array}
\right)$$
\hrulefill
\end{figure*}

$(24,12,8)$ extended Golay code is a quasi-perfect self-dual binary linear block code  \cite{codetheoryEng}. One of many possible ways to describe it is given by the Turyn construction  \cite{assmus1967research}.
The codewords are obtained as  \[c=(u+v,u+w,u+v+w),v,w\in C',u\in C'',\]
where  $C'$  is the $(8,4,4)$ extended Hamming code, and $C''$ is a code equivalent to  $C$, such that  $C'\cap C''=\set{(0,0,0,0,0,0,0,0),(1,1,1,1,1,1,1,1)}$. Note that both $C'$ and $C''$ are instances of extended BCH codes with generator polynomials $g'(x)=(x-\alpha)(x-\alpha^2)(x-\alpha^4)=x^3+x+1$ and $g''(x)=(x-\alpha^{3})(x-\alpha^6)(x-\alpha^5)=x^3+x^2+1$, where $\alpha$ is a primitive element of $\F_{2^3}$.

Their generator and check matrices are given by 
$$G'=H'=\begin{blockarray}{cccccccc}
0 & \alpha^0 & \alpha^1&\alpha^2&\alpha^3 & \alpha^4 & \alpha^5&\alpha^6 \\
\begin{block}{(cccccccc)}
1&1&1&0&1&0&0&0\\
1&0&1&1&0&1&0&0\\
1&0&0&1&1&0&1&0\\
1&0&0&0&1&1&0&1\\
\end{block}
\end{blockarray}
$$
and 
$$G''=H''=\begin{blockarray}{cccccccc}
0 & \alpha^0 & \alpha^1&\alpha^2&\alpha^3 & \alpha^4 & \alpha^5&\alpha^6 \\
\begin{block}{(cccccccc)}
1&1&0&1&1&0&0&0\\
1&0&1&0&1&1&0&0\\
1&0&0&1&0&1&1&0\\
1&0&0&0&1&0&1&1\\
\end{block}
\end{blockarray}.
$$
The columns of the matrices are indexed with elements of $\F_{2^3}$. Arranging these elements in the standard bit order $(0,1,\alpha,\alpha+1=\alpha^3,\alpha^2,\alpha^2+1=\alpha^6,\alpha^2+\alpha=\alpha^4,\alpha^2+\alpha+1=\alpha^5)$, and combining the matrices according to the Turyn construction, one obtains the generator and check matrix for the extended Golay code shown at the top of this page.

In order to employ the successive cancellation algorithm and its derivatives for decoding of the extended Golay code, we define a mixed polarizing transformation  $A=\diag(A_4,A_3)$. Then, for a suitable matrix $Q$, one obtains the constraint matrix $V$ shown at the top of this  page.

\section{Decoding}
\label{sDecoding}
\subsection{Chained decoding schedule}
\label{sChainedSchedule}
Decoding of the extended Golay code in the above proposed chained representation can be implemented using two instances of the Tal-Vardy list decoder, which are configured for polarizing transformations $A_4$ and $A_3$, respectively. Each instance is responsible for memory management, path cloning and computing path probabilities or LLRs. However, these instances need to be synchronized. The synchronization is achieved by computing the global path score \eqref{mScoreMinSum}, where the log-likelihood ratios $S_m^{(i)}$ are computed by either of the corresponding Tal-Vardy decoder instances. 

According to the greedy procedure given the \cite{trifonov2017chained},  one obtains the following sequence of symbol $u_i$ indices to be processed by the decoder: $0,1,2,16,3,17,4,5,18,6,7,8,9,19,20, 10,21,11,12,22,13$, $14,15,23.$
For example, the initial four steps of decoding according to this schedule correspond to frozen symbols $u_0=u_1=u_2=u_{16}=0$.
Hence, one obtains  a single all-zero path with the score 
\begin{align*}
\mathcal A=R(0000|y_0^{23})=&\tau(S_4^{(0)}(y_0^{15}),0)+\tau(S_4^{(1)}(0,y_0^{15}),0)\\
&+\tau(S_4^{(2)}(00,y_0^{15}),0)+\tau(S_3^{(0)}(y_{16}^{23}),0).
\end{align*}
Then one needs to consider two possible values of $u_3$, i.e. clone the path. This immediately enables one to process freezing constraint $u_{17}=u_3$, which follows from the equation $u_0^{23}V^T=0$.
Hence, one obtains 
\begin{align*}
R(0000u_3u_{17}|y_0^{23})=&\mathcal  A+\tau(S_4^{(3)}(000,y_0^{15}),u_3)\\
&+\tau(S_3^{(1)}(y_{16}^{23}),u_{17}).
\end{align*}
The decoder operates in the same way until paths of length $24$ are obtained.
The result of decoding is given by the path with the highest score.  
The decoding complexity can be substantially reduced by employing the sequential  algorithm described in \cite{miloslavskaya2014sequential}.
\subsection{Block decoding }
\label{sBlockDecoding}
The decoding complexity can be  reduced by joint processing of some blocks of the input symbols of the polarizing transformation \cite{trofimiuk2015block}. In order to exploit this approach, we observe that puncturing last 8 symbols transforms the extended Golay code into $(16,11,4)$ extended Hamming code. It can be represented as a Plotkin concatenation of the $(8,4,4)$ first-order Reed-Muller code, and a single-parity check code. 
Observe also, that puncturing all codeword symbols for the extended Golay code except those with indices $16,\dots,23$ results in $(8,7,2)$ single parity check code, which can be obtained via Plotkin concatenation of the $(4,3,2)$ first-order Reed-Muller code and $(4,4,1)$ trivial code.  Rows 6,7 of matrix $V$ provide linear relations between the codewords of $(8,4,4)$ and $(4,3,2)$ codes.

The correlation metrics for the codewords of a first-order Reed-Muller code $\mathcal C$ of length $N-1$
$$\mathbf C(c^{(i)},z_0^{N-1})=\sum_{j=0}^N (-1)^{c_j^{(i)}}z_i,c^{(i)}\in\mathcal C, $$
where $z_i$ are the log-likelihood ratios, 
can be obtained via order-$N$ fast Hadamard transform (FHT) with complexity $N\log_2 N$ summations.  Given a correlation metric, the corresponding ellipsoidal weight can be computed as $$E(c^{(i)},z_0^{N-1})=\frac{1}{2}\left(\sum_{j=0}^{N-1} |z_j|-\mathbf C(c^{(i)},z_0^{N-1})\right).$$
This implies that 
\begin{equation}
2R(u_0^7,u_{16}^{19}|y_0^{23})=-\sum_{j=0}^{11} |z_j| +C(u_0^7A_3,z_0^{7})+\mathbf  C(u_{16}^{19}A_2,z_8^{11}),
\end{equation}
where  $z_i=S_1^{(0)}(y_{2i},y_{2i+1})= S_{0}^{(0)}(y_{2i})\boxplus S_{0}^{(0)}(y_{2i+1}), 0\leq i<12,$ and $u_0=u_1=u_2=u_4=u_{16}=0$, $u_3=u_{17}$, $u_5=u_{18}$. Observe that the first summand does not depend on $u_0^{23}$, and can be neglected. With this simplification, one obtains 
$R(u_0^6,u_7=1,u_{16}^{18},u_{19}=1|y_0^{23})=-R(u_0^6,u_7=0,u_{16}^{18},u_{19}=0|y_0^{23})$.
Hence, the scores of 32 paths $(u_0^7,u_{16}^{19})$ can be computed via order-8 and order-4 FHTs and 16 additional summations. We propose to sort these pathes in the  descending order\footnote{Observe that only $16$ values need to be actually sorted.}, and apply the below described second processing step until a stopping condition is satisfied.

For any path $(u_0^7,u_{16}^{19})$ with score $r=2R(u_0^7,u_{16}^{19}|y_0^{23})$ one can compute $u_9=u_{20}=u_3+u_5+u_6+u_{19}$.
Now one can compute $\widetilde z_{8+i}=S_1^{(1)}(u_{16}^{19}A_3,y_{16+2i}^{16+2i+1}), 0\leq i<3$. These can be considered as the LLRs for a codeword of the coset, given by the value of $u_{20}$, of $(4,3,2)$  code . Hence, one can compute the corresponding correlation metrics using the order-4 FHT and obtain scores  $$\rho=2R(u_0^7,u_{16}^{23}|y_0^{23})=r-\sum_{i=8}^{11}|\widetilde z_i|+\mathbf C(u_{20}^{23}A_3,\widetilde z_8^{11}).$$ 
Note that only vectors with $\mathbf C(u_{20}^{23}A_3,\widetilde z_8^{11})\geq 0$ need to be considered, since $u_{23}$ is not frozen. 
Let the vectors $u_{20}^{23}$ be ordered in the descending order of $\mathbf C(u_{20}^{23}A_3,\widetilde z_8^{11})$.  
Now one can compute $u_{10}=u_3+u_5+u_{21}$ and $u_5=u_{22}$.
Let us further compute 
$\widetilde z_{i}=S_1^{(1)}(u_{0}^{7}A_4,y_{2i}^{2i+1}), 0\leq i<7$. The can be considered as the LLRs for a coset, given by $u_9.u_{10},u_{12}$, of the $(8,4,4)$ first order Reed-Muller code. Hence, one can use order-8 FHT to compute the correlation metrics, and finally select the codeword with the highest value of 
$$2R(u_0^{23}|y_0^{23})=r-\sum_{i=0}^{11}|\widetilde z_i|+\mathbf C(u_{20}^{23}A_3,\widetilde z_8^{11})+\mathbf C(u_{8}^{15}A_4,\widetilde z_0^{7}).$$
Observe that coefficients $2$ and $1/2$ in the above equations can be omitted. 
 
In order to avoid redundant calculations, one should keep the highest value $R_{max}$ of $R(u_0^{23}|y_0^{23})$ obtained so far, and abort processing of vectors $u_{20}^{23}$ as soon as  one obtains the value of $\rho<R_{max}$, and abort processing of $(u_0^7,u_{16}^{19})$ as soon as one obtains $r<R_{max}$.

The best-case complexity of the above described algorithm corresponds to the case when the correct codeword has the highest values of $\mathbf C(u_0^7A_3,z_0^{7})+\mathbf  C(u_{16}^{19}A_2,z_8^{11})$ and $\mathbf C(u_{20}^{23}A_3,\widetilde z_8^{11})$, and  exactly two FHTs of order 4 and 3 are computed. In this case the algorithm requires  $111$ summations and $45$ comparisons. 

At high signal-to-noise ratios one can further reduce the best-case decoding complexity by constructing the hard-decision vector for $\widetilde z_0^{11}$ corresponding to a given path $(u_0^7,u_{16}^{19})$, and computing the  values of $u_{20}^{22}$. If the obtained vector satisfies the constraints given by matrix $V$, one can skip computing FHTs in the second step of the algorithm.

\section{Numeric results}
\label{sNumeric}
\begin{figure}
\includegraphics[width=0.5\textwidth]{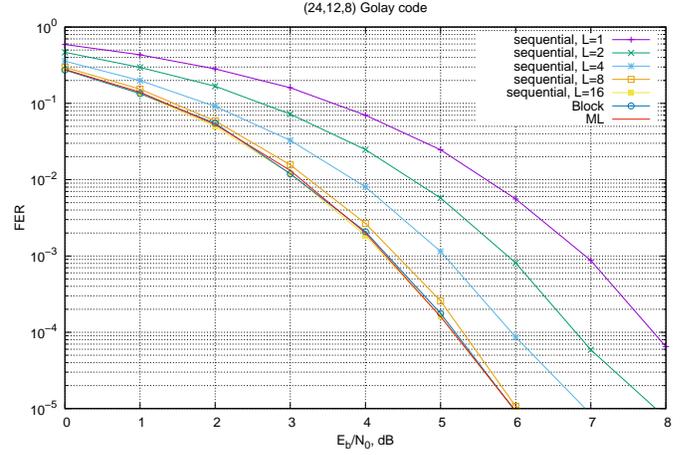}
\caption{Performance of the proposed decoding algorithms}
\label{fPerf}
\end{figure}

Figure \ref{fPerf} illustrates the performance of the extended Golay code
for the case of AWGN\ channel with BPSK\ modulation. We consider sequential decoding \cite{miloslavskaya2014sequential}  using the schedule presented in Section  \ref{sChainedSchedule}, and the block algorithm introduced in Section \ref{sBlockDecoding}. It can be seen that sequential decoding with $L=16$ provides maximum likelihood decoding. This is the expected result, since the proposed decoding schedule requires one to process four unfrozen symbols ($u_3,u_5,u_6,u_7$), before one can process all freezing constraints which involve these symbols. Hence, one needs list size at least $16$ in order to avoid killing the correct path at an early phase of decoding process. It can be seen  that the proposed block algorithm also provides maximum likelihood decoding. 

\begin{figure}
\includegraphics[width=0.5\textwidth]{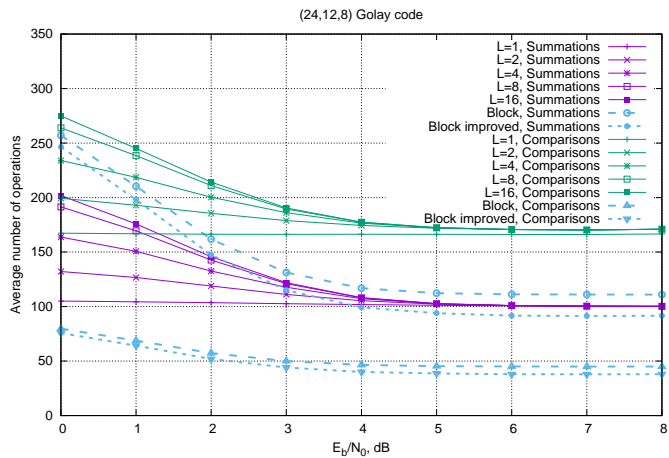}
\caption{Complexity of the proposed decoding algorithms}
\label{fComplexity}
\end{figure}
Figure \ref{fComplexity} illustrates the average number of arithmetic operations for the proposed decoding algorithms. It can be seen that their complexity quickly decreases with SNR. At high SNR it approaches the complexity of the most efficient decoding algorithm for the Golay code \cite{vardy1995even}, which requires 121 operations. The improved block decoding algorithm, which employs hard decisions to avoid computing FHTs at the second step, provides approximately 20\% complexity reduction.

 The maximal complexity of the block algorithm observed in our simulations was 1590 operations, which is close to the complexity of the  FHT-based decoding algorithm suggested in \cite{beery1986optimal}. 

\section{Conclusions}
It was shown in this paper that the extended Golay code can be represented like a chained polar subcode. This enables one to decode it using the successive cancellation decoding algorithm and its list/sequential generalizations. With appropriate parameter selection, these algorithms can provide maximum likelihood decoding.  The decoding complexity can be reduced by exploiting the fast Hadamard transform.

Although the complexity of these algorithms is slightly higher than the complexity of the Vardy algorithm, which was designed specifically for the extended Golay code, the proposed approach enables one to decode this code using the same techniques as polar codes. Since polar codes were recently adopted for use in 5G, many communication systems are likely to have an implementation of a decoder for polar codes. The proposed approach enables one to reuse the corresponding hardware, and avoid implementing dedicated circuitry for decoder the extended Golay code, reducing thus the overall implementation complexity. It remains an open problem to identify other types of error-correcting codes, which can be decoded in the same way.

A similar representation of the extended Golay code as a punctured twisted polar code was independently derived in \cite{bioglio2018polar}. However, the authors considered only the straightforward implementation of the successive cancellation list decoder.

\end{document}